\documentclass[preprint,prc,aps,showpacs,showkeys,groupedaddress,floatfix]{revtex4}
\usepackage{epsfig}
\usepackage{dcolumn}
\usepackage{bm}
\usepackage[latin5]{inputenc}
\usepackage{graphics}
\usepackage{graphicx}
\usepackage{epsfig}
\usepackage{amssymb}
\usepackage{amsmath}
\begin{document}
\title{Scattering, bound and quasi-bound states of the generalized symmetric Woods-Saxon potential}
\author{B.C. L\"{u}tf\"{u}o\u{g}lu, F. Akdeniz and O. Bayrak}
\affiliation{Department of Physics, Akdeniz University, 07058
Antalya, Turkey}
\date{\today}
\begin{abstract}
The exact analytical solutions of the Schr\"{o}dinger equation for
the generalized symmetrical Woods-Saxon potential are examined for
the scattering, bound and quasi-bound states in one dimension. The
reflection and transmission coefficients are analytically obtained.
Then, the correlations between the potential parameters and the
reflection-transmission coefficients are investigated, and a
transmission resonance condition is derived. Occurrence of the transmission
resonance has been shown when incident energy of the particle is equal
to one of the resonance energies of the quasi-bound states.
\end{abstract}
\keywords{Generalized symmetric Woods-Saxon potential, Scattering
states, bound states, quasi-bound states, analytical solutions}
\pacs{03.65.Ge} \maketitle 
\section{Introduction}
The investigation of the transmission resonance of a quantum
particle has raised a great deal of interest in relativistic and
non-relativistic quantum mechanics in the last two decades
\cite{Dombey,Kennedy1,Kennedy2,VillalbaWSKG,alpdogan,nurayhulthen,VillalbaDiracsymmetriccup,VillalbaDiracsymmetriccup2,VillalbaDiracdoublebarrier,diracmass1,diracmass2},
especially, the supercriticality concept which has a transmission
resonance at zero momentum has been studied for scattering particles
through a barrier potential that has a half bound for the inverted
version of the barrier potential   \cite{Dombey,Kennedy1,Kennedy2}.
Dombey \emph{et. al} have shown that scattering of the Dirac
particles through a square barrier potential lead to a transmission
resonance at zero momentum. In other words their reflection and
transmission coefficients are zero and one, respectively
\cite{Dombey}.

On the other hand, it has been shown that the Woods-Saxon potential plays an important role in atom-molecule \cite{atom} and nuclear physics for
both scattering and bound states \cite{Woods,bound-scat,scattering}.  In the literature, we find out that the transmission resonance and the supercriticality have been examined for a particle tunneling through Wood-Saxon potential \cite{Kennedy1,Kennedy2}. There, they have solved the Dirac equation in which the particle has a half-bound state  for potential well at  $E=-m$ and an anti-particle for the potential barrier at $E=m$. Furthermore, scattering and bound
state solutions of the one-dimensional mass dependent Dirac equation with the Woods-Saxon potential have been obtained and supercriticality conditions for different mass functions have been presented \cite{diracmass1,diracmass2}. Not only Woods-Saxon potential, but also other potentials such as asymmetric Hulth\'{e}n potential \cite{nurayhulthen}, symmetric screened potential \cite{VillalbaDiracsymmetriccup,VillalbaDiracsymmetriccup2},
a double-cusp barrier \cite{VillalbaDiracdoublebarrier} have been investigated.

The transmission resonance and supercriticality concept have also been examined for a Klein-Gordon particle moving under the Woods-Saxon potential \cite{VillalbaWSKG}. Rojas and  Villalba have been obtained the transmission resonance  for arbitrary potential parameters and have been concluded that there is no supercritical states. Moreover, solution of the Klein-Gordon equation for the asymmetric Woods-Saxon potential has been examined for both scattering and bound states. The transmission resonance and supercriticaly conditions are given in Ref. \cite{alpdogan}.

In literature, not only scattering states, also bound states solutions have been studied for
the Woods-Saxon potential and its modified versions for the Schr\"{o}dinger \cite{EckartWS,fluge,bizws} and the Dirac equations \cite{diracws}.
Furthermore, within the framework of the spin and pseudospin symmetry, the analytical solutions of the Woods-Saxon and its generalized versions are solved for Dirac equation and obtained the bound state energies with  their corresponding wave functions for the particle and antiparticle \cite{pseudo1,nurayspin}.

In this paper, we propose a potential model that is the symmetric
version of the generalized Woods-Saxon potential \cite{bizws}. The
generalized symmetric Woods-Saxon (GSWS) potential model is more
flexible and useful model than the Woods-Saxon potential in order to
examine the scattering, bound and quasi-bound state solutions of the
wave equations so that the GSWS potential can be applied to physical
phenomena such as the scattering, transmission resonance,
supercriticality, decay, fusion, fission \emph{etc.}. Especially,
the GSWS potential can be useful model in description of the surface
interaction of the
particles\cite{surface1,surface2,surface3,surface4,surface5}. Since
GSWS potential model has many applications in physics, here we show
how to solve the Schr\"{o}dinger equation analytically for the GSWS
potential in one dimension in case of the scattering, bound and
quasi-bound states. Moreover, we examine correlations between the
potential parameters with the reflection-transmission coefficients
in the case of scattering state, with the energy eigenvalues and
their corresponding wave functions in the case of bound state, and
finally with the resonance energy eigenvalues and their
corresponding wave functions in the case of quasi-bound states.

In the following section, we present the analytical solution of the
Schr\"{o}dinger equation in one dimension for the GSWS potential model for the following
cases: scattering, bound and quasi-bound states. We discuss the analytic and numeric results of the model
potential. In section \ref{conclusion}, our conclusion is given.

\section{The Model}
\begin{figure}[!htb]
\centering
\includegraphics[totalheight=0.5\textheight]{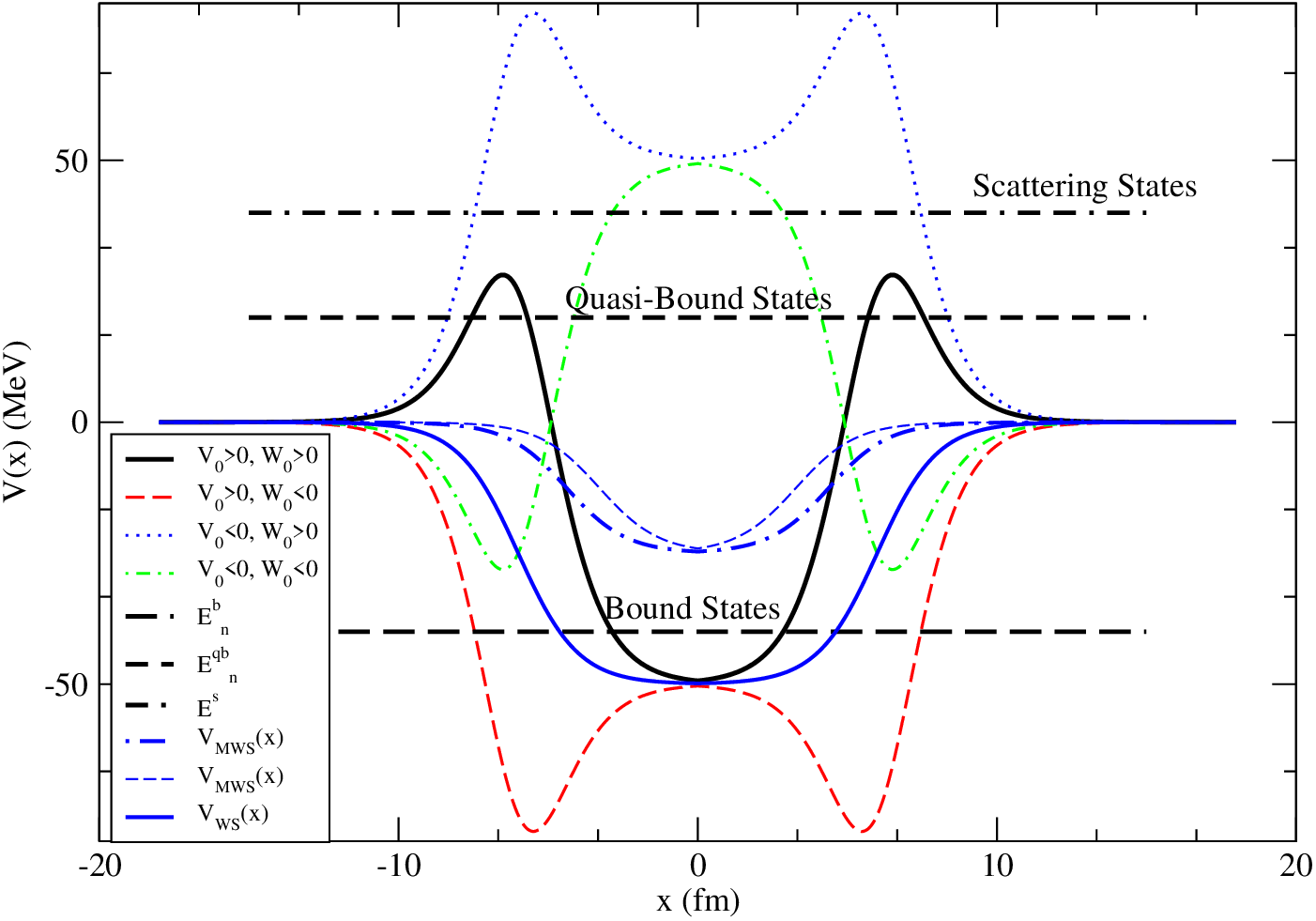}
   \caption{Shapes of the various GSWS and MWS potentials for different signs of the potential depth parameters. The magnitudes of the potential parameters are $V_0=50$ $MeV$, $W_0=200$ $MeV$, $a=1$ $fm^{-1}$ and
   $L=6$ $fm$. The particle energies are assumed as $E^b_n=-40$ $MeV$, $E^{qb}_n=20$ $MeV$ and
   $E^s=40$ $MeV$. In figure we present the bound, quasi-bound and scattering states of a particle for $V_0>0$ and $W_0>0$. The GSWS potential reduces the Woods-Saxon potential for $W=0$, $V_{WS}(x)$, (blue solid line).
   The blue dot-dashed line shows  MWS potential $V_{MWS}(x)$ for $p=2$ and $q=10$ and the blue dashed line shows  MWS potential $V_{MWS}(x)$ for $p=2$ and $q=30$.} \label{pot}
\end{figure}

The GSWS potential in one dimension can be given by,
\begin{eqnarray}\label{gws}
  V(x)&=&\theta{(-x)}\Bigg[-\frac{V_0}{1+e^{-a(x+L)}}+\frac{W_0 e^{-a(x+L)}}{\big(1+e^{-a(x+L)}\big)^2}\Bigg] + \theta{(x)}\Bigg[-\frac{V_0}{1+e^{a(x-L)}}+\frac{W_0
  e^{a(x-L)}}{\big(1+e^{a(x-L)}\big)^2}\Bigg], \nonumber\\
 \end{eqnarray}
where $\theta{(\pm x)}$ are the Heaviside step functions. The depth parameters
of the potential $V_0$ and $W_0$ can be positive or negative.
In the model there are other positive and real parameters $L$ and $a$, namely control parameters, adjust the shape of the potential.
In Fig.(\ref{pot}) we present various shapes of the GSWS potential versus changing the sings of the potential depth parameters.

GSWS potential given in Eq.(\ref{gws}) differs from the Woods-Saxon
potential by its second terms in the square brackets. While $V_0>0$,
these terms modify the potential at the surface region which
constitute either a pocket for  $W_0<0$, or a barrier for $W_0>0$.
This provides a great advantage in description of the interacting
particle for the bound, quasi-bound and scattering states
\cite{surface1,surface2,surface3,surface4,surface5}. Hence the model
potential is very useful to determine the behavior of a particle in
the bound, quasi-bound and scattering states. In the literature
there is also the modified version of Woods-Saxon potential(MWS)
similar to following equation \cite{alpdogan}, 
\begin{eqnarray}\label{mws}
  V(x)&=&\theta{(-x)}\Bigg[-\frac{V_0}{p+qe^{-a(x+L)}}\Bigg] + \theta{(x)}\Bigg[-\frac{V_0}{p+qe^{a(x-L)}}\Bigg],
 \end{eqnarray}
where $p>0$ and $q>0$ are integer numbers. In Fig.(\ref{pot}), we
also show the WS, $V_{WS}(x)$ and MWS, $V_{MWS}(x)$ potentials. The
MWS potential reduces to WS potential for $p=1$ and $q=1$ in
Fig.(\ref{pot}). When increasing $p$ parameter, depth of the
potential decreases for MWS potential. As increasing $q$, geometry
of the potential changes both volume and surface regions of the
potential. The GSWS potential have two advantages. Firstly, we can
modify any region of the Woods-Saxon potential with surface term.
The surface term of the potential has significant important in
description of surface
interaction\cite{surface1,surface2,surface3,surface4,surface5}.
Secondly, we can simultaneously examine interaction of the particle
with GSWS potential for the bound, scattering and quasi-bound states
in terms of the convenient potential parameters in Fig.(\ref{pot}).

In present paper we only focus on the case of $V_0>0$ and $W_0>0$
depth parameters and examine,
\begin{itemize}
  \item the scattering states $E^s>0$,
  \item the bound states $-V_0<E^b_{n}<0$,
  \item the quasi-bound states $0<E^{qb}_n<\frac{(V_0-W_0)^2}{4W_0}$.
\end{itemize}
Here $E^b_n,E^{qb}_n$ and $E^s$ are the energies of the particle in
the bound, quasi-bound and scattering states, respectively. The term
$\frac{(V_0-W_0)^2}{4W_0}$ is height of the barrier (HB) which can
be derived from the potential. One can choose different signs of the
potential depth parameters and  examine other shapes of potential
that is represented in Fig.(\ref{pot}) the scattering $E^s>0$ states
in case of $V_0>0$ and $W_0<0$, one can only investigate bound and
scattering states since there is no barrier term in the potential.
Moreover, in case of $V_0<0$ and $W_0>0$, one can study the
quasi-bound and scattering states. Finally, in case of $V_0<0$
$W_0<0$, one can examine the bound, quasi-bound and scattering
states.

A final remark before we proceed, the model potential is completely symmetric with respect to y-axis
\emph{i.e} $V(-x)=V(x)$. Therefore we have to obtain even and odd solutions for the GSWS potential.

We consider a particle with a mass of $m$ moving under the GSWS
potential, one dimensional Scr\"{o}dinger equation can be given by,
\begin{equation}\label{sch}
    \Big[\frac{d^2}{dx^2}+\frac{2m}{\hbar^2}\Big(E-V(x)\Big)\Big]\phi(x)= 0.
\end{equation}
Due to discontinuity in the potential, we examine the analytical
solution of the GSWS potential at two regions, i.e, $x<0$ and $x>0$.
Inserting Eq.(\ref{gws}) into Eq.(\ref{sch}) for $x<0$ region, we
have,
\begin{eqnarray} \label{Schx<0}
  \Bigg[\frac{d^2}{dx^2}+\frac{2m}{\hbar^2}\Bigg(E-\frac{W_0-V_0}{1+e^{-a(x+L)}}+\frac{W_0}{\big(1+e^{-a(x+L)}\big)^2}\Bigg)\Bigg]\phi_L(x) &=& 0.
\end{eqnarray}
By mapping $z\equiv\Big[1+e^{-a(x+L)}\Big]^{-1}$ and using the
following descriptions, 
\begin{eqnarray} \label{definition}
  -\epsilon^2\equiv\frac{2m E}{a^2\hbar^2},\,\,\ \beta^2\equiv\frac{2m(V_0-W_0)}{a^2\hbar^2},\,\,\  \gamma^2\equiv\frac{2m
  W_0}{a^2\hbar^2},
\end{eqnarray}
we obtain,
\begin{eqnarray} \label{Sch1x<0}
  \Bigg[\frac{d^2}{dz^2}+\frac{2z-1}{z(z-1)}\frac{d}{dz}+\frac{1}{z^2 (z-1)^2}\bigg(-\epsilon^2+\beta^2 z+ \gamma^2
  z^2\bigg)\Bigg]
  \phi_L(z) &=& 0.
\end{eqnarray}
In this equation, there are two singular points, \emph{i.e.}, $z=0$ and
$z=1$. In order to remove these singularities, we have to examine
asymptotic behavior of Eq.(\ref{Sch1x<0}). At $x\rightarrow -\infty$
($z\rightarrow 0$) limit, the dominant terms in
Eq.(\ref{Sch1x<0}) are, 
\begin{eqnarray} \label{Schasm1x<0}
    \Bigg[\frac{d^2}{dz^2}+\frac{1}{z}\frac{d}{dz}-\frac{\epsilon^2}{z^2}\Bigg]\phi_L(z) &=& 0.
\end{eqnarray}
To solve this equation we define $\phi_L(z)\equiv z^\mu$ with
$\mu\equiv\epsilon\equiv\frac{ik}{a}$ and $k=\sqrt{\frac{2m E}{\hbar^2}}$.

On the other hand, at $x\rightarrow 0$ ($z\rightarrow 1$) limit, the
dominant terms in Eq.(\ref{Sch1x<0}) are, 
  \begin{eqnarray} \label{Schasm2x<0}
    \Bigg[\frac{d^2}{dz^2}+\frac{1}{z-1}\frac{d}{dz} +\frac{\beta^2+\gamma^2-\epsilon^2}{(z-1)^2}\Bigg]\phi_L(z) &=& 0.
  \end{eqnarray}
We define $\phi_L(z)\equiv(z-1)^\nu$ with
$\nu\equiv\frac{i\kappa}{a}$ and
$\kappa\equiv\sqrt{\frac{2m}{\hbar^2}(E+V_0)}$. Therefore we suggest the
wavefunction as $\phi_L(z)\equiv z^\mu (z-1)^{\nu}f(z)$ and insert into
Eq.(\ref{Sch1x<0}) we obtain,
\begin{eqnarray}\label{2f1x<0}
   z(1-z)f{''}+ \Big[(1+2\mu)-(1+2\mu+2\nu+1)z\Big]f'-\Big[2\mu \nu+\mu+\nu+2\epsilon^2-\beta^2\Big]f&=&
   0.
\end{eqnarray}
We compare Eq.(\ref{2f1x<0})  with the hypergeometric
equation \cite{Abramowitz} which is, 
\begin{eqnarray}
   z(1-z)w{''}+ \Big[c-(1+a+b)z\Big]w'- ab w &=& 0,
\end{eqnarray}
and its analytical solution is,
\begin{eqnarray} \label{hipergeometrik_genel_cozum}
  w(z) &=& A\,\,\, {}_2F_1[a,b,c;z]+B z^{1-c}\,\,\, {}_2F_1[1+a-c,1+b-c,2-c;z].
\end{eqnarray}
We can easily obtain the general solution of the GSWS potential for
$x<0$ region as follows:
\begin{equation}\label{leftwave}
  \phi_L(z) = D_1 z^\mu (z-1)^{\nu} \,\,\, {}_2F_1[a_1,b_1,c_1;z]+D_2 z^{-\mu} (z-1)^{\nu} \,\,\,
  {}_2F_1[1+a_1-c_1,1+b_1-c_1,2-c_1;z],
\end{equation}
where $a_1$, $b_1$ and $c_1$ are,
\begin{eqnarray} \label{abcL}
  a_1 &\equiv& \mu+\theta+\nu, \nonumber\\
  b_1 &\equiv&1+\mu-\theta+\nu, \nonumber\\
  c_1 &\equiv&1+2\mu.
\end{eqnarray}
Here $\theta\equiv\frac{1}{2}\mp\sqrt{\frac{1}{4}-\gamma^2}$.

In case of $x>0$, the one dimensional Schr\"{o}dinger equation for
the GSWS potential in Eq.(\ref{gws}) becomes,
\begin{eqnarray}\label{Sch1x>0}
  \Bigg[\frac{d^2}{dx^2}+\frac{2m}{\hbar^2}\Bigg(E-\frac{W_0-V_0}{1+e^{a(x-L)}}+\frac{W_0}{\big(1+e^{a(x-L)}\big)^2}
  \Bigg)\Bigg]\phi_R(x) &=& 0.
\end{eqnarray}
By using the transformation $t\equiv\Big[1+e^{a(x-L)}\Big]^{-1}$ and the
definitions in Eq.(\ref{definition}) we easily get, 
\begin{equation}
  \Bigg[\frac{d^2}{dt^2}+\frac{2t-1}{t(t-1)}\frac{d}{dt}+\frac{1}{t^2 (t-1)^2}\bigg(-\epsilon^2+\beta^2 t+ \gamma^2 t^2\bigg)\Bigg]
  \phi_R(t) = 0.
\end{equation}
By using the procedure after Eq.(\ref{Sch1x<0}), in terms of
Eq.(\ref{abcL}) we can obtain,
\begin{equation}\label{rightwave}
  \phi_R(t) = D_3 t^\mu (t-1)^{\nu} \,\,\, {}_2F_1[a_1,b_1,c_1;t]+D_4 t^{-\mu} (t-1)^{\nu} \,\,\,
  {}_2F_1[1+a_1-c_1,1+b_1-c_1,2-c_1;t].
\end{equation}
After obtaining the wave function of the GSWS potential, we can
examine the interaction of the particle in the GSWS potential for
the case of the continuum, bound and quasi-bound states.

\subsection{The Continuum States}\label{continuum}
In the continuum states, we assume that the particle is coming from
negative infinity and going to positive infinity. Additionally, we could also assume the
particle as incident from the right side of the  potential well, due
to symmetric form of the potential we could obtain the same results that we find from left side. In the continuum states the energy of the particle is
positive ($E^s>0$ and $k>0$) and has continuum values. In all
calculations, we assume $aL>>1$ which adjusts the width of the barrier. In case of continuum states the
wave functions are given by Eq.(\ref{leftwave}) and
Eq.(\ref{rightwave}) with $\mu=\frac{ik}{a}$ and $k=\sqrt{\frac{2m
E^s}{\hbar^2}}$ for $k^2>0$, $\nu=\frac{i\kappa}{a}$ and
$\kappa=\sqrt{\frac{2m}{\hbar^2}(E^s+V_0)}$. In the negative domain
of $x$-axis, we have to investigate the asymptotic behavior of the
wave function in Eq.(\ref{leftwave}). In case of
$x\rightarrow$-$\infty$ ($z\rightarrow 0$), $z\approx e^{a(x+L)}$
and $_2F_1[a,b,c;0]=1$. Therefore we have the incident and reflected
waves, 
\begin{eqnarray}\label{leftwaveAsymptoticInfinity}
  \phi_L(x\rightarrow -\infty,z\rightarrow 0) \rightarrow e^{-\frac{\pi \kappa}{a}} \Big(D_1 e^{ik(x+L)}+D_2 e^{-ik(x+L)}
  \Big).
\end{eqnarray}
In case of $x\rightarrow0$ ($z\rightarrow 1$), we have to
investigate the behavior of the hypergeometric function by using the
relation \cite{Abramowitz}, 
\begin{eqnarray}\label{HypergeoZ=1}
  _2F_1(a,b,c;y) &=& \frac{\Gamma(c)\Gamma(c-a-b)}{\Gamma(c-a)\Gamma(c-b)} \,\,\, _2F_1(a,b, a+b-c+1;1-y)\nonumber \\
  &&+(1-y)^{c-a-b}\frac{\Gamma(c)\Gamma(a+b-c)}{\Gamma(a)\Gamma(b)}\,\,\, _2F_1(c-a,c-b,
  c-a-b+1;1-y),\,\,\,\,
\end{eqnarray}
and considering $(1-z)\approx e^{-a(x+L)}$ for $z\rightarrow 1$ we
get the wave function at the vicinity $z=1$ as,
\begin{eqnarray}\label{leftwaveAsymptoticZero}
 \phi_L(x\rightarrow 0,z\rightarrow 1)\rightarrow \Big[(D_1 N_1+D_2N_3) e^{-i\kappa(x+L)}
 + (D_1 N_2+D_2 N_4 ) e^{i\kappa(x+L)} \Big]e^{-\frac{\pi
 \kappa}{a}},
\end{eqnarray}
in terms of the following definitions:
\begin{eqnarray}\label{leftwaveDefinitions}\nonumber
  N_1 &\equiv& \frac{\Gamma(c_1)\Gamma(c_1-a_1-b_1)}{\Gamma(c_1-a_1)\Gamma(c_1-b_1)}, \\ \nonumber
  N_2 &\equiv& \frac{\Gamma(c_1)\Gamma(a_1+b_1-c_1)}{\Gamma(a_1)\Gamma(b_1)}, \\  \nonumber
  N_3 &\equiv& \frac{\Gamma(2-c_1)\Gamma(c_1-a_1-b_1)}{\Gamma(1-a_1)\Gamma(1-b_1)}, \\
  N_4 &\equiv&
  \frac{\Gamma(2-c_1)\Gamma(a_1+b_1-c_1)}{\Gamma(1+a_1-c_1)\Gamma(1+b_1-c_1)}.
\end{eqnarray}
At the region $x>0$, let's examine the behavior of the wave function
in Eq.(\ref{rightwave}) for the $(x\rightarrow$$0,t\rightarrow 1)$
and $(x\rightarrow$$\infty,t\rightarrow 0)$ cases. In case of
$x\rightarrow0$ ($t\rightarrow1$), using Eq.(\ref{HypergeoZ=1})
and considering $(1-t)\approx e^{a(x-L)}$ we have,
\begin{eqnarray}\label{rightwaveAsymptoticZero}
  \phi_R(x\rightarrow 0) \rightarrow \Big[(D_3  M_1 +D_4 M_3)e^{i\kappa(x-L)} + (D_3 M_2+D_4M_4) e^{-i\kappa(x-L)} \Big]e^{-\frac{\pi \kappa}{a}},
\end{eqnarray}
with the following relations:
\begin{eqnarray}\label{rightwaveDefinitions}
  M_1=N_1, M_2=N_2, M_3=N_3, M_4=N_4.
\end{eqnarray}
In case of $x\rightarrow$$\infty$ ($t\rightarrow 0$), $t\approx
e^{-a(x-L)}$ and $_2F_1[a,b,c;0]=1$. As a result we obtain the
reflected and transmitted waves as follows: 
\begin{eqnarray}\label{rightwaveAsymptoticInfinity}
  \phi_R(x\rightarrow \infty,t\rightarrow 0) \rightarrow \Big[D_3 e^{-ik(x-L)} (-1)^{\nu}+D_4 e^{ik(x-L)} \Big]e^{-\frac{\pi \kappa}{a}}.
\end{eqnarray}
Since the potential at the infinity is not definded, the term $D_3$ should
be zero and we have only transmitted waves for $x\rightarrow\infty$.

\begin{figure}[!htb]
\centering
\includegraphics[totalheight=0.55\textheight]{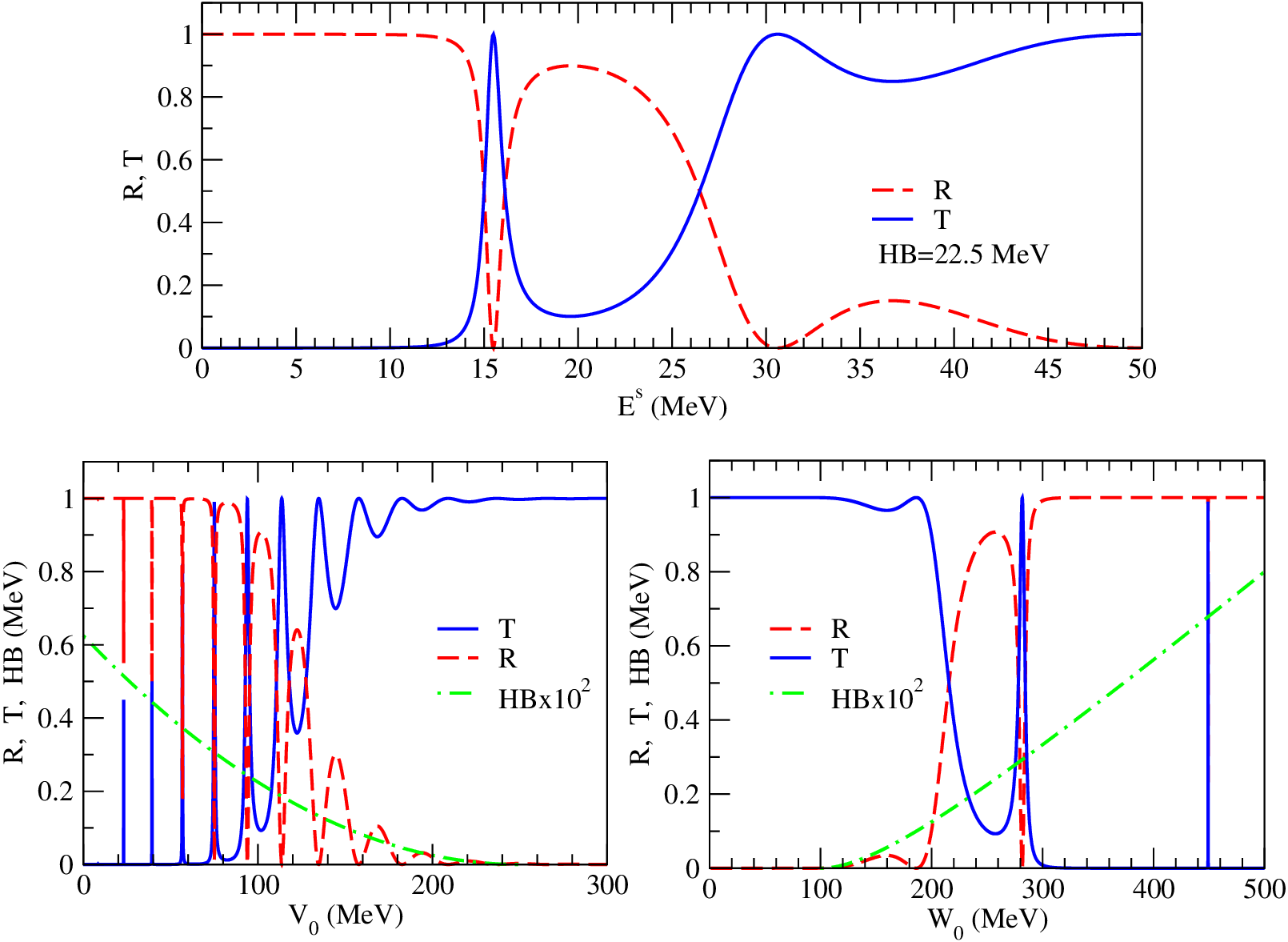}
   \caption{Variations of the reflection, transmission coefficients and the HB as a function of $E^s$, $V_0$ and $W_0$ parameters.
   Here we take $V_0=100$ $MeV$, $W_0=250$ $MeV$, $L=6$ $fm$ and $a=1$ $fm^{-1}$ for variation of R and T with $E^s$ on top panel. Other parameters are assumed constant,
   we take $E^s=20$ $MeV$ for variation of the $V_0$ and $W_0$ parameters. In calculations we use the following values: $mc^2=940$ $MeV$ and $\hbar c= 197.329$ $MeV.fm$.} \label{RT}
\end{figure}

\begin{figure}[!htb]
\centering
\includegraphics[totalheight=0.60\textheight]{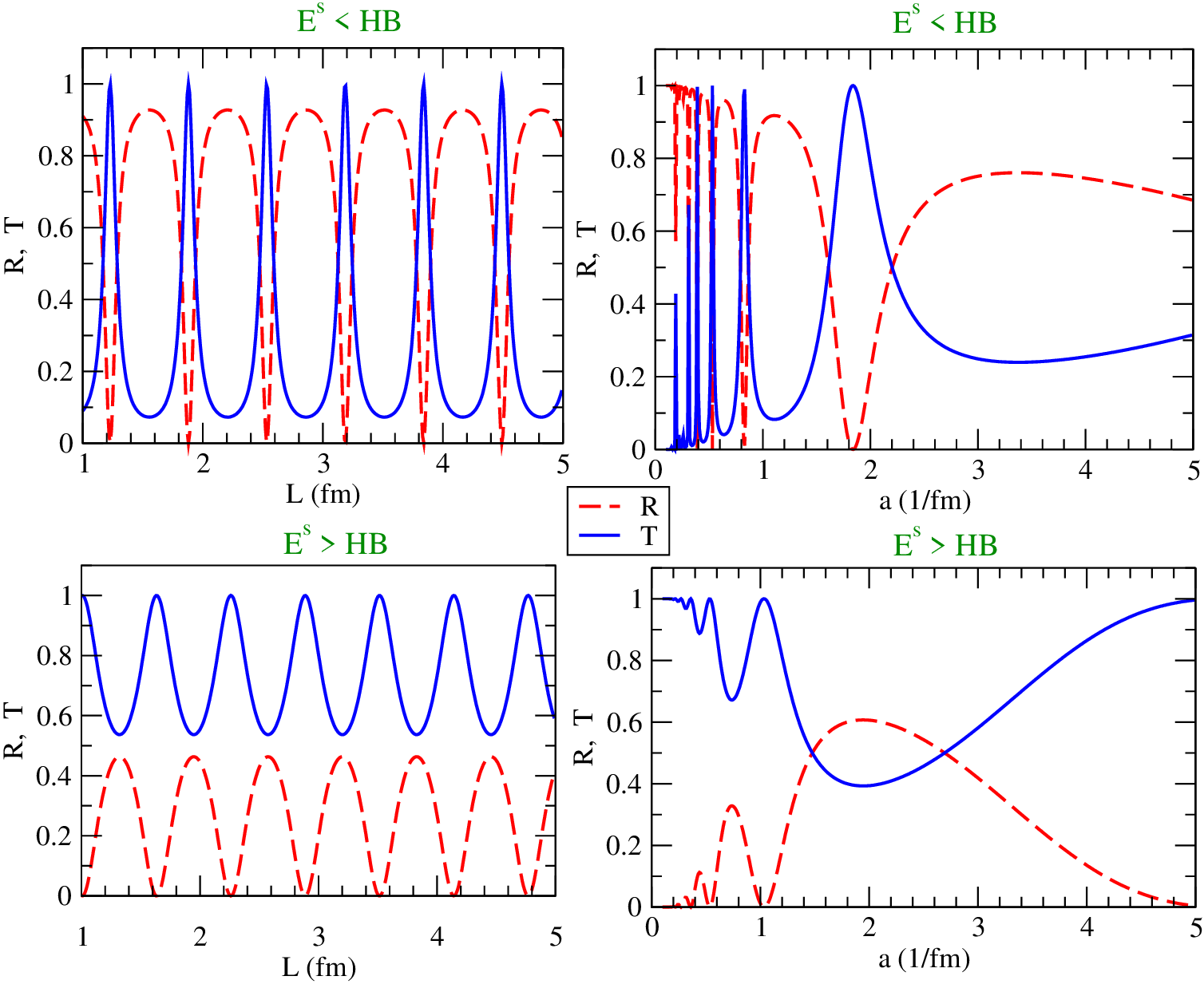}
   \caption{Variation of the reflection and transmission coefficients versus potential control parameters $a$ and $L$ in cases of $E^s<HB$ and $E^s>HB$.
   Here we take $V_0=100$ $MeV$, $W_0=250$ $MeV$, $mc^2=940$ $MeV$ and $\hbar c= 197.329$ $MeV.fm$. On top panels, we present variation of $R$ and $T$ depending on $a$ and $L$    while the particle has lower energy than height of the barrier namely, $E^s=20$ $MeV$$<$$HB=22.5$ $MeV$ (Quantum tunneling). On bottom panels, we  plot the variation of $R$ and $T$ depending on $a$ and $L$ while the particle has bigger energy than height of the barrier \emph{i.e}, $E^s=30$ $MeV$ $>$ $HB=22.5$ $MeV$ (Resonance scattering).} \label{RTal}
\end{figure}

Since our model potential has a discontinuity due to the Heaviside
step functions, we have to use the continuity conditions: At $x=0$,
$\phi_L(x=0)=\phi_R(x=0)$ and
$\frac{d}{dx}\phi_L(x=0)=\frac{d}{dx}\phi_R(x=0)$ should be
satisfied. Therefore we get, 
\begin{eqnarray}\label{continuumroot}
  D_1 N_1+D_2N_3-D_4 M_3  &=&  (D_4M_4 -D_1 N_2-D_2 N_4)e^{2i\kappa L}, \\ \nonumber
    D_1 N_1+D_2N_3+D_4 M_3  &=&  (D_4M_4 -D_1 N_2+D_2 N_4)e^{2i\kappa
    L}.
\end{eqnarray}
Taking $N_i=M_i$ and $i=1,2,3,4$ in Eqs. (\ref{leftwaveDefinitions})
and (\ref{rightwaveDefinitions}) into account, and solving
Eq.(\ref{continuumroot}), we obtain 
\begin{eqnarray}
  \frac{D_2}{D_1}&=&\frac{\Big[\frac{N_1}{N_4}e^{-2i\kappa L}-\frac{N_2}{N_3}e^{2i\kappa L}\Big]}{\Big[\frac{ N_4}{N_3 }e^{2i\kappa L}-\frac{N_3}{N_4}e^{-2i\kappa L}\Big]}, \\
   \frac{D_4}{D_1}  &=&\frac{\frac{N_1}{N_3}-\frac{N_2}{N_4}}{\Big[\frac{ N_4}{N_3 }e^{2i\kappa L}-\frac{N_3}{N_4}e^{-2i\kappa
   L}\Big]}.
\end{eqnarray}
It is known that the reflected and transmitted coefficients are
defined by $R=\frac{J_{refl.}}{J_{inc.}}$ and
$T=\frac{J_{trans.}}{J_{inc.}}$. Here $J_{inc.}$, $J_{refl.}$ and
$J_{trans.}$ are the incident, reflected and transmitted probability
currents and obtained by using,
\begin{eqnarray}\label{flux}
  j(x,t)&=&\frac{\hbar}{2mi}\Big[\phi^*\frac{d\phi}{dx}-\phi\frac{d\phi^*}{dx}\Big],
\end{eqnarray}
which satisfies the continuity equation given by
\begin{eqnarray}\label{continuityEq}
  \frac{\partial j(x,t)}{\partial x}+\frac{\partial\rho(x,t)}{\partial
  t}&=&0,
\end{eqnarray}
where the probability density function
$\rho(x,t)=\phi(x)^*\phi(x)=|\phi(x)|^2$. The reflection and
transmission coefficients of the wave function with Eq.(\ref{flux})
are obtained as $R=\Big|\frac{D_2}{D_1}\Big|^2$ and $T=\Big|\frac{D_4}{D_1}\Big|^2$.
Therefore we can easily get,
\begin{eqnarray}\label{reflect}
  R &=& \frac{2-(\frac{N_1N_3}{N_2N_4}e^{-4i\kappa L}+\frac{N_2N_4}{N_1N_3}e^{4i\kappa L})}{\frac{N_1N_4}{N_2N_3}+\frac{N_2N_3}{N_1N_4}
  -\Big[\frac{N_2N_4}{N_1N_3}e^{4i\kappa L}+\frac{N_1N_3}{N_2N_4}e^{-4i\kappa
  L}\Big]},
\end{eqnarray}
and
\begin{eqnarray}\label{trans}
  T &=& \frac{\frac{N_1N_4}{N_2N_3}+\frac{N_2N_3}{N_1N_4}-2}{\frac{N_1N_4}{N_2N_3}+\frac{N_2N_3}{N_1N_4}
  -\Big[\frac{N_2N_4}{N_1N_3}e^{4i\kappa L}+\frac{N_1N_3}{N_2N_4}e^{-4i\kappa
  L}\Big]}.
\end{eqnarray}
In order to take into account $N_1^*=N_4$ and $N_2^*=N_3$, we can
analytically obtain $R+T=1$. We would like to construct  a condition
for transmission resonance \emph{i.e.,} $T=1$ and
$R=0$ \cite{Kennedy1}. In order to satisfy the resonance
transmission in the transition coefficient equation,
Eq.(\ref{trans}), the square-bracketed term in the denominator
should be equal to two. Therefore we have,
\begin{eqnarray}\label{transcondition}
  \sin(4\kappa
  L)=-\frac{i}{2}\Bigg[\frac{(N_1N_3)^2-(N_2N_4)^2}{N_1N_2N_3N_4}\Bigg].
\end{eqnarray}
It can be seen that the resonance condition depends on the
potential parameters ($V_0, W_0, L, a$) and incident energy of the
particle ($E^s$). In Fig.(\ref{RT}), the reflection and
transmission coefficients are plotted as a function of incident particle energy
$E^s$ and the depth parameters $V_0$ and $W_0$. The HB of the potential is $22.5$ $MeV$ for the parameters
$V_0=100$ $MeV$, $W_0=250$ $MeV$, $L=6$ $fm$ and $a=1$ $fm^{-1}$. Therefore, as the incident
energy of the particle is very low, there is a total reflection
($R=1$ and $T=0$) in Fig.(\ref{RT}). In calculation we cannot obtain
any transmission resonance for low energy or at zero energy for any
potential parameters. Increasing incident energy of the particle,
the resonances begin to be observed in Fig.(\ref{RT}). At the
resonance energy, the reflection and transmission coefficients have
extremum values and their minimum and maximum values are $0$ and $1$,
respectively. First resonance energy is $E^s=15.4913$ $MeV$ which can
be obtained using Eq.(\ref{transcondition}). Other resonance
energies are $30.6153$ $MeV$ and $50.37$ $MeV$. In case of $E^s>>HB$,
increasing the incident energy of the particle, the reflection and
transmission coefficients become $0$ and $1$, respectively, in
Fig.(\ref{RT}). On the bottom of Fig.(\ref{RT}) we present
dependencies of the reflection and transmission coefficients on
depths of the potential $V_0$ and $W_0$. The HB
changes with $V_0$ and $W_0$. We also plot variation of the HB with
$V_0$ and $W_0$ parameters in Fig.(\ref{RT})(bottom panel). In order
to have the HB, it should be noted that it is
$W_0>V_0$. This case is explicitly shown in Fig.(\ref{RT}). The
maximum and minimum of the reflection and transmission coefficients
depending on $V_0$ and $W_0$ parameters are apparently seen in
Fig.(\ref{RT}). These maximum and minimum can also be calculated by
using Eq.(\ref{transcondition}). For very small $V_0$ values the
reflection and transmission coefficients are $1$ and $0$, respectively.
But decreasing of height of the barrier with increasing $V_0$, the
reflection and transmission coefficients go to $0$ and $1$, respectively,
in Fig.(\ref{RT}) (left-bottom panel). At the minimum of the HB, the
reflection and transmission coefficients are $0$ and $1$ as expected.
For very small values of $W_0$ the reflection and transmission
coefficients are $0$ and $1$ respectively since there is a very small
barrier in Fig.(\ref{RT}) (right-bottom panel). As the HB increase with
increasing $W_0$, the reflection and transmission coefficients go to
$1$ and $0$, respectively, in Fig.(\ref{RT}) (right-bottom panel). In
Fig.(\ref{RTal}) we also plot variation of the reflection and
transmission coefficients as a function of $a$ and $L$ parameters in
case of the incident energies are lower than the HB ($E^s<HB$) and
higher than the HB ($E^s>HB$). The maximum and minimum values of the
reflection and transmission coefficients can be calculated by using
Eq.(\ref{transcondition}) for variation of  $a$ and $L$ parameters.

\subsection{The Bound States}\label{bound}
In the bound states, the particle is inside the potential well and
the energy of the particle is quantized as well as $E^b_n<0$.
Therefore we have $k_n\equiv\sqrt{-\frac{2mE^b_n}{\hbar^2}}$ and
$\kappa_n \equiv \sqrt{\frac{2m}{\hbar^2}(E^b_n+V_0)}$. At the $x<0$ and
$x>0$ regions, the wave functions are equal to Eq.(\ref{leftwave}) and
Eq.(\ref{rightwave}) respectively, with $\mu \equiv \frac{k_n}{a}$ and
$\nu \equiv \frac{i\kappa_n}{a}$.

Let's examine the asymptotic behavior of
the wave function at $x<0$ region in Eq.(\ref{leftwave}) for $x\rightarrow -\infty$ ($z\rightarrow0$). In this limit  case
$z\approx e^{a(x+L)}$.
Therefore we have, 
\begin{eqnarray}\label{boundsolasym}
  \phi_L(x\rightarrow -\infty,z\rightarrow 0)\rightarrow D_1 e^{k_n(x+L)} (e^{i\pi \nu}) +D_2 e^{-k_n(x+L)} (e^{i\pi
  \nu}).
\end{eqnarray}
In case of bound states, the wave function should be zero for
$x\rightarrow-\infty$  ($z\rightarrow 0$) in
Eq.(\ref{boundsolasym}). Therefore second term in
Eq.(\ref{boundsolasym}) should be zero in order to satisfy the
boundary condition at $x<0$, \emph{i.e.}, $D_2=0$. As a result, the
wave function at $x<0$ is
\begin{eqnarray}
\phi_L(z)= D_1 z^\mu (z-1)^{\nu} \,\,\,{}_2F_1(a,b,c;z).
\end{eqnarray}
 For the case of $x\rightarrow0$  ($z\rightarrow
1$), we have to investigate the behavior of the hypergeometric
function by using the relation Eq.(\ref{HypergeoZ=1}) and taking
$(1-z)\approx e^{-a(x+L)}$, we get,
\begin{eqnarray}\label{boundsol1}
 \phi_L(x\rightarrow 0,z\rightarrow 1) \rightarrow \Big[D_1 N_1 e^{-i\kappa_n(x+L)} + D_1 N_2 e^{i\kappa_n(x+L)} \Big]e^{-\frac{\pi
 \kappa_n}{a}},
\end{eqnarray}
where $ N_1$ and $ N_2$ are given in Eq.(\ref{leftwaveDefinitions}).
At $x>0$ region, we have to examine asymptotic behavior of the wave
function for bound states in Eq.(\ref{rightwave}). In case of
$x\rightarrow0$ ($t\rightarrow 1$), by using
Eq.$(\ref{HypergeoZ=1})$ and taking $(1-t)\approx e^{a(x-L)}$ into
account, we obtain, 
\begin{eqnarray}\label{boundsag1}
  \phi_R(x\rightarrow0,t\rightarrow 1)\rightarrow \Big[(D_3  M_1 +D_4 M_3)e^{i\kappa_n(x-L)} + (D_3 M_2+D_4M_4) e^{-i\kappa_n(x-L)} \Big]e^{-\frac{\pi \kappa_n}{a}},
\end{eqnarray}
where $M_1$, $M_2$, $M_3$ and $M_4$ are given in
Eq.(\ref{rightwaveDefinitions}). For the case of $x\rightarrow
+\infty$  ($t\rightarrow0$), considering $t\approx e^{-a(x-L)}$ and
$_2F_1(a,b,c;0)=1$, we have 
\begin{eqnarray}\label{boundsagasym}
  \phi_R(x\rightarrow \infty,t\rightarrow 0)\rightarrow \Big[D_3 e^{-k_n(x-L)} (-1)^{\nu}+D_4 e^{k_n(x-L)} \Big]e^{-\frac{\pi \kappa_n}{a}}.
\end{eqnarray}
In this equation second term does not satisfy the boundary condition
\emph{i.e.}, $\phi_R(x\rightarrow \infty,t\rightarrow 0)=0$.
Therefore $D_4$ should be zero in Eq.(\ref{boundsagasym}).

The left $\phi_L(x)$ and right $\phi_R(x)$ wave functions should be
continuous  at $x=0$. Before we apply the continuity condition, we
present the latest form of the left $\phi_L(x)$ and right $\phi_R(x)$ wave
functions. Taking $D_4=0$ in Eq. (\ref{boundsag1}) and $N_1=M_1$ and
$N_2=M_2$ in  Eq.(\ref{rightwaveDefinitions}), we have
\begin{eqnarray}
 \phi_L(x\rightarrow0)\rightarrow \Big[D_1 N_1 e^{-i\kappa_n(x+L)}
 + D_1 N_2 e^{i\kappa_n(x+L)} \Big]e^{-\frac{\pi \kappa_n}{a}}, \nonumber \\
 \phi_R(x\rightarrow0) \rightarrow \Big[D_3  N_1 e^{i\kappa_n(x-L)} + D_3 N_2 e^{-i\kappa_n(x-L)} \Big]e^{-\frac{\pi \kappa_n}{a}}.
\end{eqnarray}
By using the continuity conditions $\phi_L(0)=\phi_R(0)$, we get 
\begin{eqnarray}\label{boundeven}
   (D_1-D_3)\big[N_1e^{-i\kappa_nL}+ N_2e^{i\kappa_nL}\big]&=& 0,
\end{eqnarray}
and using $\frac{d}{dx}\phi_L(x=0)=\frac{d}{dx}\phi_R(x=0)$ we
obtain,
\begin{eqnarray}\label{boundodd}
  (D_1+ D_3)\big[N_1e^{-i\kappa_nL}- N_2e^{i\kappa_nL}\big]&=& 0.
\end{eqnarray}
The GSWS potential in one dimension has a symmetry under the space
transformations $x\rightarrow-x$. Therefore we have even and odd
solution for Eq.(\ref{boundeven}) and Eq.(\ref{boundodd}).

\begin{figure}[!htb]
\centering
\includegraphics[totalheight=0.55\textheight]{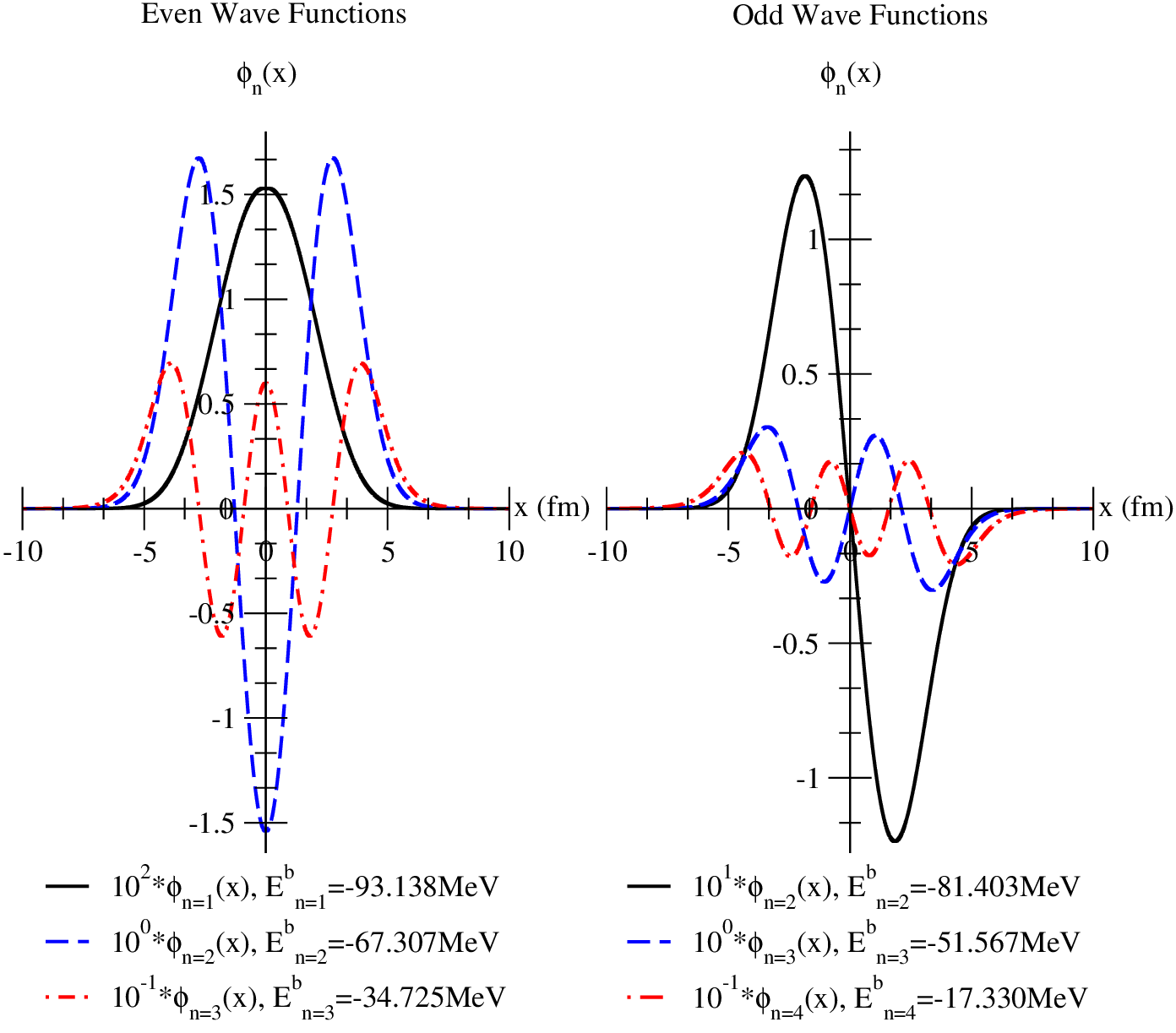}
   \caption{Plot of the even and odd unnormalized wave functions for some energy eigenvalues for the bound states.
   In calculations  we take $V_0=100$ $MeV$, $W_0=250$ $MeV$, $L=6$ $fm$, $a=1$ $fm^{-1}$, $mc^2=940$ $MeV$ and $\hbar c= 197.329$ $MeV.fm$.} \label{plotwaves}
\end{figure}

\textbf{Even Solution:} In Eq.(\ref{boundeven}), taking $D_1=D_3$ we
obtain the bound state energy eigenvalue equation as,
\begin{eqnarray}\label{boundenergyeven}
 E^b_n&=&-V_0+\frac{\hbar^2}{2mL^2}\Bigg[\arctan{\bigg[\frac{(N_1-N_2)}{i(N_1+N_2)}\bigg]}\pm{n\pi}\Bigg]^2,
 n=0,1,2,...
\end{eqnarray}
and the corresponding wave function at $x<0$ as,
\begin{eqnarray}\label{boundwaveevenleft}
\phi_L(x)  &=& D_1 \frac{e^{\frac{k_n-i\kappa_n}{2}(x+L)-\frac{\pi
\kappa_n}{a}}}{\big[2\cosh\big(\frac{a}{2}(x+L)\big)\big]^{\frac{k_n+i\kappa_n}{a}}}
                         \,\,\,
                         {}_2F_1\Big[a_1,b_1,c_1;\frac{1}{1+e^{-a(x+L)}}\Big],
\end{eqnarray}
and at $x>0$ region as,
\begin{eqnarray}\label{boundwaveevenright}
  \phi_R(x) &=&D_1     \frac{e^{\frac{-k_n+i\kappa_n}{2}(x-L)-\frac{\pi \kappa_n}{a}} }{\big[2\cosh\big(\frac{a}{2}(x-L)\big)\big]^{\frac{k_n+i\kappa_n}{a}}}
             \,\,\,
             {}_2F_1\Big[a_1,b_1,c_1;\frac{1}{1+e^{a(x-L)}}\Big].\nonumber\\
\end{eqnarray}
\textbf{Odd Solution:} In Eq.(\ref{boundodd}), taking $D_1=-D_3$ we
obtain the bound state energy eigenvalue equation as,
\begin{eqnarray}\label{boundenergyodd}
E^b_n&=&-V_0+\frac{\hbar^2}{2mL^2}\Bigg[\arctan{\bigg[\frac{(N_1+N_2)}{i(N_1-N_2)}\bigg]}\pm{n\pi}\Bigg]^2,
n=0,1,2,...
\end{eqnarray}
and the corresponding wave function at $x<0$ region as,
\begin{eqnarray}\label{boundwaveoddleft}
  \phi_L(x) &=& D_1     \frac{e^{\frac{k_n-i\kappa_n}{2}(x+L)-\frac{\pi \kappa_n}{a}} }{\big[2\cosh\big(\frac{a}{2}(x+L)\big)\big]^{\frac{k_n+i\kappa_n}{a}}}
             \,\,\,
             {}_2F_1\Big[a_1,b_1,c_1;\frac{1}{1+e^{-a(x+L)}}\Big],
\end{eqnarray}
and at $x>0$ region as,
\begin{eqnarray}\label{boundwaveoddright}
  \phi_R(x) &=&-D_1     \frac{e^{\frac{-k_n+i\kappa_n}{2}(x-L)-\frac{\pi \kappa_n}{a}} }{\big[2\cosh\big(\frac{a}{2}(x-L)\big)\big]^{\frac{k_n+i\kappa_n}{a}}}
             \,\,\,
             {}_2F_1\Big[a_1,b_1,c_1;\frac{1}{1+e^{a(x-L)}}\Big].
\end{eqnarray}
In Fig.(\ref{plotwaves}), we present the even and odd wave functions
of the bound states for some energy eigenvalues by using
Eqs.(\ref{boundenergyeven}, \ref{boundwaveevenleft}) and
Eq.(\ref{boundwaveevenright}) for even solutions as well as
Eqs.(\ref{boundenergyodd}, \ref{boundwaveoddleft}) and
Eq.(\ref{boundwaveoddright}) for odd solutions. We calculate the
energy eigenvalues and corresponding wave functions of the GSWS
potential for the given parameters which are $V_0=100$ $MeV$,
$W_0=250$ $MeV$, $L=6$ $fm$, $a=1$ $fm^{-1}$, $mc^2=940$ $MeV$ and
$\hbar c= 197.329$ $MeV.fm$. In these parameters we can only obtain
four energy eigenvalues $E^b_{n=1}=-93.138$ $MeV$,
$E^b_{n=2}=-67.307$ $MeV$, $E^b_{n=3}=-34.725$ $MeV$,
$E^b_{n=4}=-0.125$ $MeV$ which satisfy the boundary conditions for
even solutions, respectively. Similarly we show that only three
energy eigenvalues can be obtained $E^b_{n=2}=-81.403$ $MeV$,
$E^b_{n=3}=-51.567$ $MeV$, $E^b_{n=4}=-17.330$ $MeV$ which satisfy
the boundary conditions for odd solutions, respectively. The number
of eigenvalues depends on the parameters, namely $V_0$, $W_0$, $L$
and $a$. One can find different eigenvalues with different
parameters that we choose. The higher quantum numbers $n$ do not
satisfy the boundary condition of the bound states due to the
positive energy. In order to obtain more solution for higher quantum
numbers we have to determine boundary condition of the particle for
$E^b_n>0$, namely we investigate quasi-bound state solution of the
GSWS potential.

\subsection{The Quasi-Bound States}\label{quasi}
In this case, the particle is inside the quantum well, but it has
positive and complex energy eigenvalues, namely $E^{qb}_n \equiv E_r-iE_i$
and $E_{r}>>E_{i}$. By using Eq.(\ref{leftwaveAsymptoticInfinity})
we determine the wave function for $x\rightarrow
-\infty$ ($z\rightarrow 0$) as 
\begin{eqnarray}\label{leftwaveAsymptoticInfinityQuasi}
  \phi_L(x\rightarrow -\infty,z\rightarrow 0)\rightarrow D_2 e^{-ik_n(x+L)}
  e^{-\frac{\pi\kappa_n}{a}},
\end{eqnarray}
where $k_n \equiv \sqrt{\frac{2m E^{qb}_n}{\hbar^2}}$ and $\kappa_n
\equiv \sqrt{\frac{2m}{\hbar^2}(E^{qb}_n+V_0)}$. In
Eq.(\ref{leftwaveAsymptoticInfinity}) we take $D_1=0$ since we have
only outgoing wave to negative infinity. For $x\rightarrow\infty$
($t\rightarrow0$) we use Eq.(\ref{rightwaveAsymptoticInfinity}) and
obtain, 
\begin{eqnarray}\label{rightwaveAsymptoticInfinityQuasi}
  \phi_R(x\rightarrow \infty,t\rightarrow 0)\rightarrow D_4 e^{ik_n(x-L)}e^{-\frac{\pi\kappa_n}{a}}.
\end{eqnarray}
In this case we have only outgoing wave that goes to positive infinity, namely
$D_3=0$. In continuum states we already obtain the behavior of the
wave functions $\phi_L(x)$ and $\phi_R(x)$ at the vicinity
$x\rightarrow 0$ ($z\rightarrow1$) and $x\rightarrow0$ ($t\rightarrow1$)
in Eq.(\ref{leftwaveAsymptoticZero}) and in
Eq.(\ref{rightwaveAsymptoticZero}). Considering $D_1=0$, $D_3=0$ and
$N_3=M_3$ as well as $N_4=M_4$, we get, 
\begin{eqnarray}
 \phi_L(x\rightarrow 0) \rightarrow \Big[(D_2N_3) e^{-i\kappa_n(x+L)}
 + (D_2 N_4 ) e^{i\kappa_n(x+L)} \Big]e^{-\frac{\pi \kappa_n}{a}}, \\
  \phi_R(x\rightarrow 0)\rightarrow \Big[(D_4 N_3) e^{i\kappa_n(x-L)} + (D_4N_4) e^{-i\kappa_n(x-L)} \Big]e^{-\frac{\pi \kappa_n}{a}}.
\end{eqnarray}
Applying the continuity conditions at $x=0$, we obtain,
\begin{eqnarray}\label{quasiboundeven}
   (D_2-D_4)\Big[N_3 e^{-i\kappa_n L}+N_4e^{i\kappa_n L}\Big]&=&0,
\end{eqnarray}
for $\phi_L(0)=\phi_L(0)$ and
\begin{eqnarray}\label{quasiboundodd}
(D_2+D_4)\Big[N_3 e^{-i\kappa_n L}-N_4e^{i\kappa_n L}\Big]&=&0.
\end{eqnarray}
for $\frac{d}{dx}\phi_L(0)=\frac{d}{dx}\phi_R(0)$. 

\begin{figure}[!htb]
\centering
\includegraphics[totalheight=0.6\textheight]{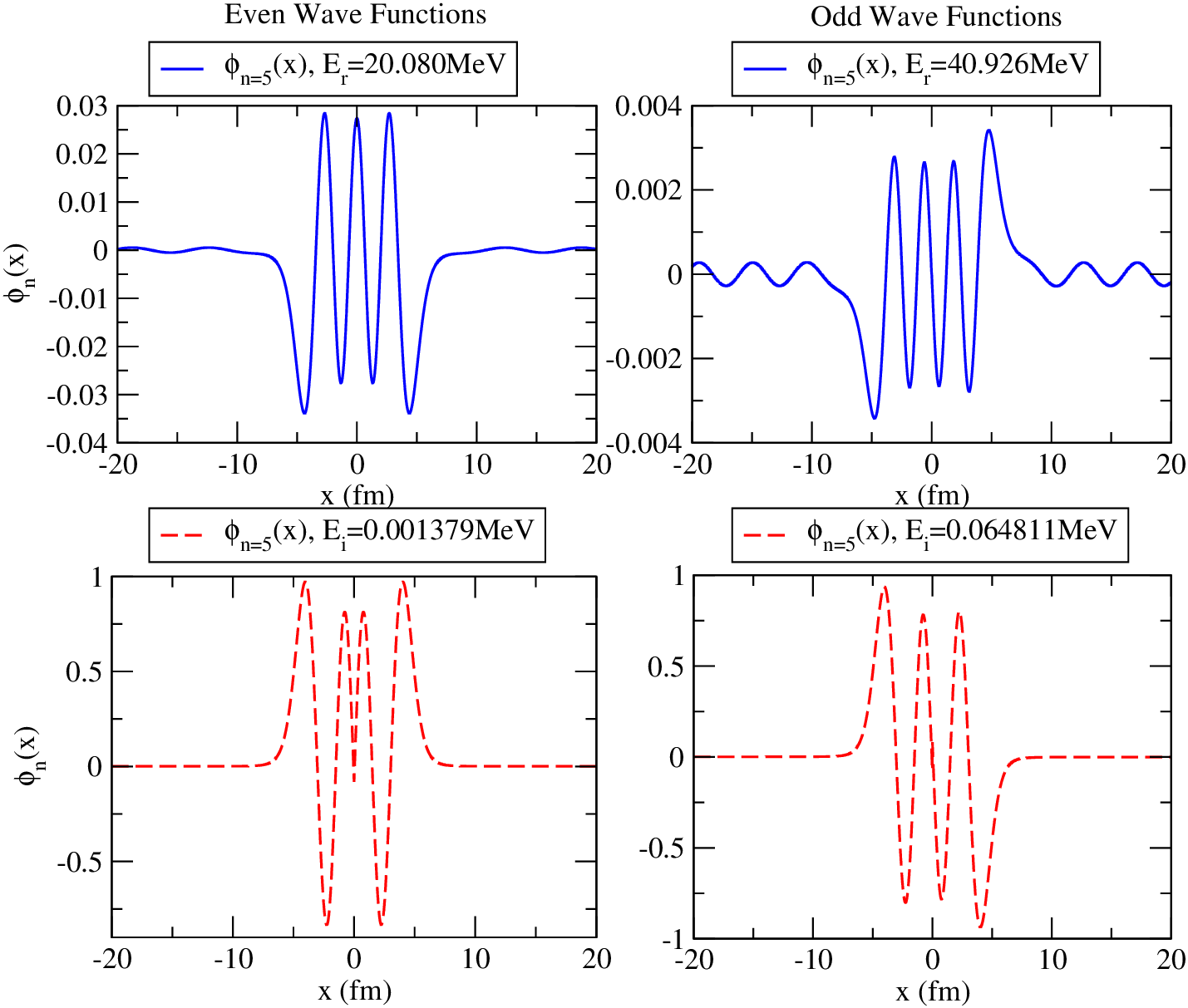}
   \caption{Plot of the even and odd unnormalized wave functions for some energy eigenvalues in case of the quasi-bound states.
   Here we take $V_0=100$ $MeV$, $W_0=450$ $MeV$, $L=6$ $fm$, $a=1$ $fm^{-1}$, $mc^2=940$ $MeV$ and $\hbar c= 197.329$ $MeV.fm$.} \label{plotquasiboundwaves}
\end{figure}

\textbf{Even Solution:} In Eq.(\ref{quasiboundeven}), taking
$D_2=D_4$ we obtain the quasi-bound state energy eigenvalue equation
as,
\begin{eqnarray}\label{quasiboundeveneigen}
 E^{qb}_n&=&-V_0+\frac{\hbar^2}{2mL^2}\Bigg[\arctan{\bigg[\frac{(N_3-N_4)}{i(N_3+N_4)}\bigg]}\pm{n\pi}\Bigg]^2,
 n=0,1,2,...
\end{eqnarray}
and the corresponding wave function at $x<0$ as,
\begin{eqnarray}\label{quasiboundevenwaveleft}
 \phi_L(x) &=& D_2 \frac{e^{\frac{-i(\kappa_n+k_n)}{2}(x+L)}e^{-\frac{\pi \kappa_n}{a}}}{\big[2\cosh\big(\frac{a}{2}(x+L)\big)\big]^{\frac{i(\kappa_n-k_n)}{a}}}
            \,\,\,{}_2F_1\Big[1+a_1-c_1,1+b_1-c_1,2-c_1;\frac{1}{1+e^{-a(x+L)}}\Big],\,\,\,\,\,\,\,\,\,\,\,\,
\end{eqnarray}
and at $x>0$ region as,
\begin{eqnarray}\label{quasiboundevenwaveright}
 \phi_R(x)&=& D_2 \frac{e^{\frac{-i(\kappa_n-k_n)}{2}(x-L)}e^{-\frac{\pi \kappa_n}{a}}}{\big[2\cosh\big(\frac{a}{2}(x-L)\big)\big]^{\frac{i(\kappa_n+k_n)}{a}}}\,\,\,
            {}_2F_1\Big[1+a_1-c_1,1+b_1-c_1,2-c_1;\frac{1}{1+e^{a(x-L)}}\Big].\,\,\,\,\,\,\,\,
\end{eqnarray}
\textbf{Odd Solution:} In Eq.(\ref{quasiboundodd}), taking
$D_2=-D_4$ we obtain the quasi-bound state energy eigenvalue
equation as,
\begin{eqnarray}\label{quasiboundoddeigen}
 E^{qb}_n&=&-V_0+\frac{\hbar^2}{2mL^2}\Bigg[\arctan{\bigg[\frac{(N_3+N_4)}{i(N_3-N_4)}\bigg]}\pm{n\pi}\Bigg]^2,
 n=0,1,2,...
\end{eqnarray}
and the corresponding wave function at $x<0$ as,
\begin{eqnarray}\label{quasiboundoddwaveleft}
 \phi_L(x) &=& D_2 \frac{e^{\frac{-i(\kappa_n+k_n)}{2}(x+L)}e^{-\frac{\pi \kappa_n}{a}}}{\big[2\cosh\big(\frac{a}{2}(x+L)\big)\big]^{\frac{i(\kappa_n-k_n)}{a}}}
            \,\,\,{}_2F_1\Big[1+a_1-c_1,1+b_1-c_1,2-c_1;\frac{1}{1+e^{-a(x+L)}}\Big],\,\,\,\,\,\,\,\,\,
\end{eqnarray}
and at $x>0$ region as,
\begin{eqnarray}\label{quasiboundoddwaveright}
 \phi_R(x)&=& -D_2 \frac{e^{\frac{-i(\kappa_n-k_n)}{2}(x-L)}e^{-\frac{\pi \kappa_n}{a}}}{\big[2\cosh\big(\frac{a}{2}(x-L)\big)\big]^{\frac{i(\kappa_n+k_n)}{a}}}\,\,\,
            {}_2F_1\Big[1+a_1-c_1,1+b_1-c_1,2-c_1;\frac{1}{1+e^{a(x-L)}}\Big].\,\,\,\,\,\,\,\,\,\,\,\,
\end{eqnarray}
The quasi-bound energy eigenvalues of the GSWS potential are same
with bound state energy eigenvalues for E$<$0. Namely, the boundary
condition of the quasi-bound state satisfies the boundary condition
of bound states. By using the potential parameters in bound state
calculation, we obtain the bound state energy eigenvalues and wave
functions in terms of Eqs.(\ref{quasiboundeveneigen},
\ref{quasiboundevenwaveleft}) and Eq.(\ref{quasiboundevenwaveright})
for even solutions and Eqs.(\ref{quasiboundoddeigen},
\ref{quasiboundoddwaveleft}) and Eq.(\ref{quasiboundoddwaveright})
for odd solutions. In case of even solutions, for $n=5$ there are no
bound and quasi-bound states since $E_{n=5}=28.6791 -i 4.24688$
$MeV$ and the particle energy has bigger energy value than the
potential barrier ($HB=22.5$ $MeV$) and scattering occurs. In case
of odd solutions, the particle has the quasi-bound energy which is
$E^{qb}_{n=5}=15.431-i0.532349$ $MeV$ for $n=5$. It should be noted
as the particle coming from left of the potential barrier has
incident energy $E^s=15.431$ $MeV$, the first resonance occurs and
the wave function of the particle is totally transmitted left side
of the potential barrier in Fig.\ref{RT}. In order to obtain
quasi-bound states for both even and odd solutions we increase depth
of the potential barrier as $W_0=450$ $MeV$ by holding other terms
constant. The energy eigenvalues of the quasi-bound states both even
and odd solutions are $E^{qb}_{n=5}=20.0801-i0.00137933$ $MeV$  and
$E^{qb}_{n=5}=40.9262-i0.0648113$ $MeV$, respectively. The
resonances occur, when the energies of the particle coming from left
of the potential barrier have incident energies as $E^s=20.0801$
$MeV$ or $E^s=40.9262$ $MeV$. We obtain these energies by using
Eq.(\ref{transcondition}). As a result, while the energies of
incident particle is equal to the quasi-bound state energy for any
$n$ quantum number, the transmission resonance occurs. In
Fig.(\ref{plotquasiboundwaves}), we show the even and odd wave
functions of the particle in case of the quasi-bound states. In
Fig.(\ref{plotquasiboundwaves}), the left and right panels show even
and odd wave functions for the real and imaginary part of the energy
eigenvalues of the particle.

\section{Conclusion}\label{conclusion}
In this paper, we have presented the exact analytical solution of
the Schr\"{o}dinger equation for the GSWS potential. We have
examined the scattering, bound and quasi-bound states of the GSWS
potential and obtained the reflection and transmission coefficients
in case of the scattering states, even and odd energy
eigenvalues and corresponding wave functions in case of bound states as well as
quasi-bound states. We have analytically shown that sum of the
reflection and transmission coefficients is equal to one. We have analytically obtained the resonance
condition and also investigated the correlations between the
reflection-transmission coefficients and the potential parameters by
considering whether the incident energy of the particle is bigger
than the potential barrier or not. Then, we have analytically obtained
the energy eigenvalues and corresponding even and odd wave functions
for bound states. Thereafter, we have considered that the particle is
inside of the potential well but has positive energy and applied the
quantum boundary conditions as well as obtained the quasi-bound
state energies of the particle and corresponding even-odd wave
functions. We have shows that while the incident energy of the particle is
equal to one of the quasi-bound state energies of the potential,
there is no reflection and particle is totally transmitted from left  to right side
of the potential barrier. However, we could not obtain transmission resonance at low or zero incident energy
of the particle for any potential parameter, namely we did not observe supercriticality for GSWS potential in non relativistic regime.

The GSWS potential model used in this paper would be useful model in order to describe
scattering states of the quantum particle, which are elastic
scattering, fusion \emph{etc.}, bound states and quasi-bound states
which are the decay of quantum particle.

\section*{Acknowledgments}
We would like to thank Dr. Esat Pehlivan and Dr. Timur Sahin for technical assistance while preparing this manuscript.
This work was supported by Research Fund of Akdeniz University. Project Number: 1031, and  partially supported by the Turkish Science and
Research Council (T\"{U}B\.{I}TAK).

\newpage
\end{document}